\documentclass{aa}

\pdfoutput=1

\usepackage[T1]{fontenc}
\usepackage[utf8]{inputenc}
\usepackage{txfonts}

\usepackage{graphicx}
\usepackage{amsmath}
\usepackage{textcomp}
\usepackage{microtype}
\usepackage{booktabs}
\usepackage{lscape}
\usepackage{tabularx}
\usepackage[inkscapelatex=false]{svg}

\usepackage[
  separate-uncertainty=true,
  range-units=single,
  range-phrase=\:\textendash\:
]{siunitx}

\DeclareSIUnit\parsec{pc}
\DeclareSIUnit\lightyear{ly}
\DeclareSIUnit\gauss{G}
\DeclareSIUnit\year{yr}
\DeclareSIUnit\years{yr}
\DeclareSIUnit\samples{samples}
\DeclareSIUnit\erg{erg}
\DeclareSIUnit\GbE{GbE}

\usepackage{color}

\usepackage[draft]{fixme}
\usepackage{soul}
\usepackage[backgroundcolor=yellow,linecolor=black]{todonotes}

\usepackage{comment}
\usepackage{xcolor}

\begin{document}

\title{The EDD Radio Astronomy Backend Framework}

\titlerunning{EDD Radio Astronomy Backend Framework}

\authorrunning{Barr et al.}

\author{
Ewan~D.~Barr\inst{1}\email{ebarr@mpifr-bonn.mpg.de}
\and
Amit~Bansod\inst{1}\email{abansod@mpifr-bonn.mpg.de}
\and
Jan~Behrend\inst{1}\email{jbehrend@mpifr-bonn.mpg.de}
\and
Weiwei~Chen\inst{1}\email{wchen@mpifr-bonn.mpg.de}
\and
Niclas~Esser\inst{1}\email{nesser@mpifr-bonn.mpg.de}
\and
Yungpen~Men\inst{1,2}\email{ypmen@hust.edu.cn}
\and
Tobias~Winchen\inst{1}\corrauth{twinchen@mpifr-bonn.mpg.de}
\and
Jason~Wu\inst{1}\email{jwu@mpifr-bonn.mpg.de}
\and
Uwe~Bach\inst{1}\email{ubach@mpifr-bonn.mpg.de}
\and
Claus~Connot\inst{1}\email{cconnot@mpifr-bonn.mpg.de}
\and
Matthias~Heininger\inst{1}\email{mhein@mpifr-bonn.mpg.de}
\and
Jedrzej~Jawor\inst{1}\email{jjawor@mpifr-bonn.mpg.de}
\and
Christoph~Kasemann\inst{1}\email{ckasemann@mpifr-bonn.mpg.de}
\and
Michael~Kramer\inst{1,3}\email{mkramer@mpifr-bonn.mpg.de}
\and
Alex~Kraus\inst{1}\email{akraus@mpifr-bonn.mpg.de}
\and
Eddy~Nussbaum\inst{1}\email{en@mpifr-bonn.mpg.de}
\and
Oliver~Polch\inst{1}\email{opolch@mpifr-bonn.mpg.de}
\and
Jan~Wagner\inst{1}\email{jwagner@mpifr-bonn.mpg.de}
\and
Gundolf~Wieching\inst{1}\email{wieching@mpifr-bonn.mpg.de}
}

\institute{
Max Planck Institute for Radio Astronomy, Auf dem H\"ugel 69, 53121 Bonn, Germany
\and
Huazhong University of Science and Technology, 1037 Luoyu Road, Wuhan 430074, China
\and
Jodrell Bank Centre for Astrophysics, The University of Manchester, Manchester M13 9PL, UK
}

\date{Received XXX; accepted YYY}

\abstract
% context heading, optional
{Modern digital radio astronomy receivers produce increasingly wide-bandwidth, high bit-rate data streams that necessitate the development of flexible, scalable, and maintainable backend processing and recording systems. Historically, such backend instrumentation has been tightly coupled to telescope observing modes, limiting reuse between observatories and science cases.}
% aims heading, mandatory
{We present the Effelsberg Direct Digitisation (EDD) backend framework, a software-defined architecture for constructing real-time radio astronomy backends on commodity off-the-shelf computing infrastructure. We describe its design, implementation, supported observing modes, and operational deployments.}
% methods heading, mandatory
{EDD separates a common core framework from plugin-provided observing capabilities. The core provides orchestration, telescope interfaces, pipeline lifecycle management, monitoring, and deployment tooling, while plugins implement processing pipelines for specific observing modes. The framework is designed to support both single-dish and interferometric instruments through site-specific configuration and plugin selection.}
% results heading, mandatory
{EDD currently supports spectroscopy and spectropolarimetry, pulsar timing and searching, baseband recording, very long baseline interferometry, correlation, and beamforming. Operational deployments include the Effelsberg 100-m telescope, the SKA-MPI prototype dish, the Thai National Radio Telescope, and the ARGOS interferometric prototype array. We show that when coupled with in-receiver digitisation, the EDD backend at Effelsberg provides $\sim30$\% improvement in pulsar time-of-arrival estimation compared with previous instruments when using the same observing bandwidth and integration time.}
% conclusions heading, optional
{EDD provides a reusable open-source framework for developing and operating real-time radio astronomy backends. By separating common services, observing-mode plugins, and site-specific configuration, it allows backend capabilities to be deployed across heterogeneous telescope environments and provides a community resource for broadband radio astronomy instrumentation.}


\maketitle
 \nolinenumbers
\section{Introduction}

Progress in astrophysics is fundamentally constrained by the quality,
bandwidth, and fidelity of observational data. In radio astronomy, this has
driven the development of receivers with increasingly large instantaneous
bandwidths, lower system temperatures, and improved stability, together with
corresponding advances in digital signal processing backends capable of
handling the resulting data volumes.

A key enabling development in modern radio astronomy systems is the transition
to early, or near-receiver, digitisation, in which the radio-frequency signal
is digitised close to the telescope focus and transported in digital form to
downstream processing systems. This approach reduces analogue signal transport
and its associated losses and stability constraints, but shifts technical
challenges to high-throughput digital data transport and processing.
Architectures of this type have been adopted in recent and next-generation
facilities, including MeerKAT and the Square Kilometre Array (SKA). These
architectures place increasing demands on downstream processing systems, making
the design of flexible and scalable backends a central challenge in modern
radio astronomy.

The evolution of such digital backends has historically been driven by the needs of specific experiments and observing modes, with early systems such as autocorrelators and digital filterbanks typically implemented on custom hardware. Subsequent generations of instruments, including spectrometers such as the XFFTS~\citep{Klein2012}, VLBI backends such as the DBBC family~\citep{Tuccari2018}, broadband interferometric backends such as CABB~\citep{CABB}, and general-purpose or multi-mode systems such as VEGAS~\citep{Prestage2015} and the GPU-based Medusa backend~\citep{Hobbs2020}, have progressively adopted programmable digital processing architectures to increase bandwidth, flexibility, and observing efficiency.

This development has been supported by reusable FPGA-based toolchains, in particular CASPER \citep{CASPER}, and hardware platforms, e.g.\ UniBoard2 \citep{uniboard2}, which have reduced the effort required to develop packetised, high-throughput signal-processing systems. More recent facility-scale instruments, including the Cobalt GPU correlator for LOFAR \citep{cobalt}, the MeerKAT correlator-beamformer~\citep{Callanan2020MeerKATXEngine, MeerKATCorrelator} and the Commensal Realtime ASKAP Fast Transient Coherent upgrade~\citep[CRACO;][]{craco}, further illustrate the increasing use of hybrid FPGA/GPU architectures and high-speed Ethernet data transport for real-time correlation, beamforming, spectral processing, and transient detection. In parallel, advances in commodity high-performance computing and networking, notably the widespread availability of 40\,GbE and 100\,GbE interconnects and the emergence of 400\,GbE and 800\,GbE technologies\footnote{\url{https://ethernetalliance.org/2025-ethernet-alliance-roadmap/}}, have enabled the use of commercial off-the-shelf~(COTS) hardware for high-data-rate, low-latency signal processing.

Despite these advances, most backend systems remain tightly coupled to specific
observing modes and instruments. Dedicated instruments are typically developed
for individual science cases, such as spectroscopy, pulsar timing, transient
searching, or VLBI, leading to a proliferation of heterogeneous systems within
and across observatories. This fragmentation increases development and
maintenance effort, complicates integration with telescope control systems, and
limits reuse of software and hardware resources. It also constrains the ability
to deploy new observing modes rapidly or to operate multiple science cases
commensally on shared infrastructure.

Early digitisation combined with COTS-based high-performance computing enables a shift away from this model. A unified, or “universal”, backend architecture can support multiple science cases within a single system, reducing duplication of effort, improving resource utilisation, and lowering the development and maintenance burden associated with independent backend systems. Such a backend must provide sufficient computational performance and scalability to process wide-band data streams in real time, support a range of observing modes through a modular and extensible design, be portable across different telescope environments, and remain maintainable over long operational timescales as scientific requirements and underlying technologies evolve. A common operational model also reduces the amount of site-specific expertise required to operate and maintain the system, facilitating shared development, remote support, and personnel rotation between instruments and observatories.

The Effelsberg Direct Digitisation (EDD) program has been developed to address
these challenges as part of a broader effort to realise modular end-to-end
observation systems for radio astronomy. In this work, we focus on the backend
component of EDD\footnote{\url{https://gitlab.mpcdf.mpg.de/mpifr-bdg/edd}}, which provides a software-defined framework for constructing
radio astronomy backends for deployment on COTS computing infrastructure. It
offers a unified architecture in which data acquisition, processing pipelines,
orchestration, and monitoring are integrated into a modular system capable of
supporting multiple science cases within a common operational framework.

In this paper, we present the design and implementation of the EDD backend and
describe how it meets the requirements of modern radio observatories. We
discuss the system architecture, including its control model, dataflow
mechanisms, and pipeline abstractions, and present representative processing
pipelines for selected science cases. We further demonstrate deployments of the
EDD backend at the Effelsberg 100-m telescope, the SKA-MPI prototype dish, and
the ARGOS interferometer prototype, illustrating its scalability and
adaptability across different operational environments.

\section{EDD Backend Design}

\subsection{System Architecture}
\label{sec:system_architecture}
\begin{figure}
\centering
\includegraphics[width=0.45\textwidth]{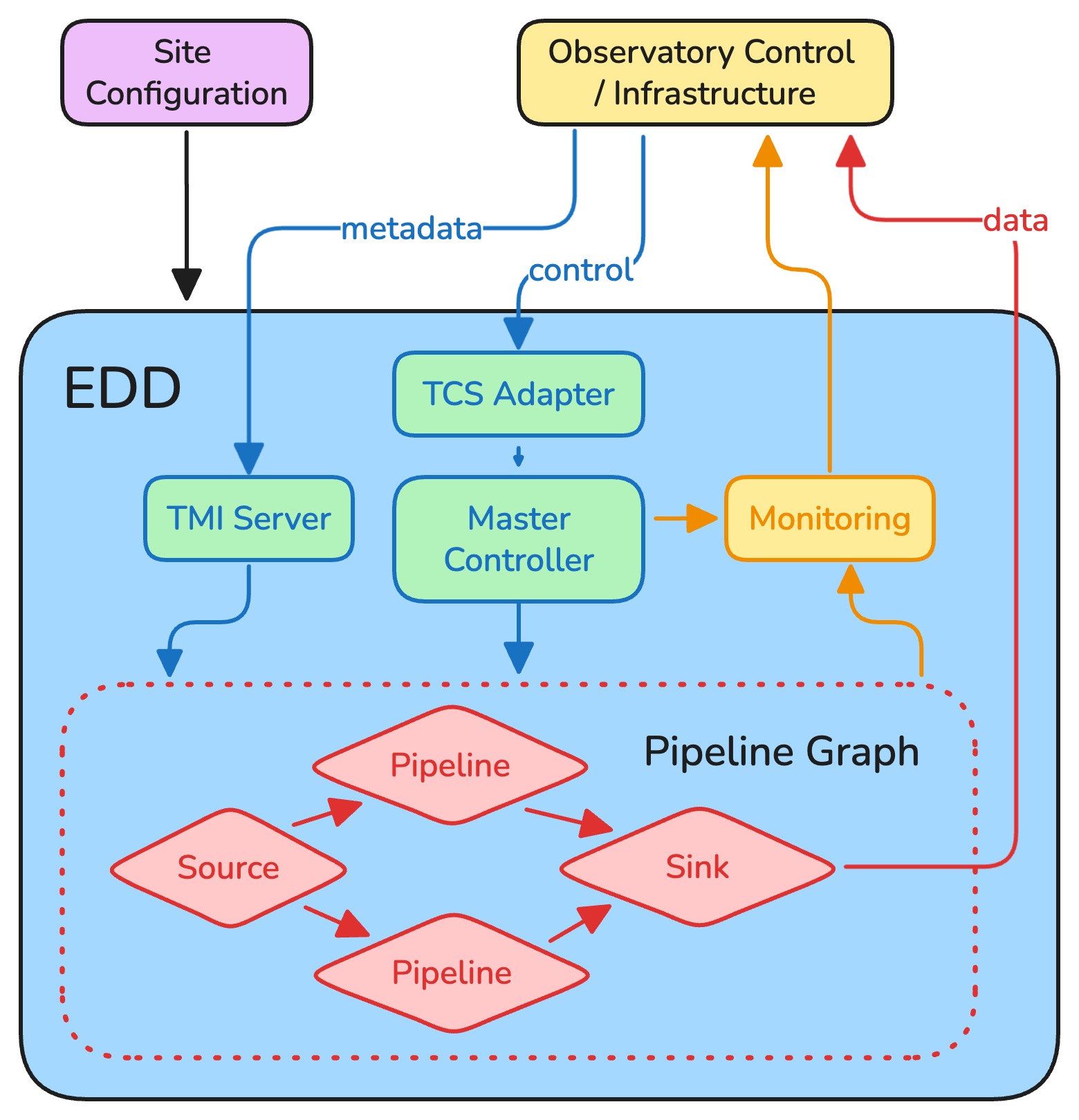}
\caption{Context diagram of relations and data flow between components in the EDD. Here TCS refers to the Telescope Control System and TMI to the Telescope Metadata Interface (see Section \ref{sec:system_architecture} for details).\label{fig:edd_concept}
}
\end{figure}

The EDD backend is designed as a modular, distributed system composed of
loosely coupled processing and control elements. At the most abstract level, it
consists of \textit{services} -- persistent components that provide
functionality such as control, interfacing, and monitoring -- and
\textit{pipelines} -- ephemeral components that execute data-processing (e.g.,
channelization) or resource-handling tasks (e.g., digitiser control). Services
are typically active for the lifetime of an EDD backend installation at a telescope site, from here on referred to as a ``deployment'', whereas pipelines
are instantiated and configured dynamically according to the observing mode and
available resources.

Pipelines are composed into observing modes by connecting them through
well-defined data streams. The resulting processing graph is a directed acyclic
graph (DAG), specified by a \textit{provision description}, in which nodes
correspond to individual pipelines and edges represent the flow of data between
them. The acyclic structure ensures a well-defined processing order and
prevents circular dependencies, enabling deterministic configuration and
execution of the system.

At each deployment site, the operation of the system is coordinated by a
\textit{Master Controller}, which manages the lifecycle of pipelines and
services, enforces the system state model, and coordinates configuration and
execution across the distributed system. All control interactions with the
backend are mediated through this component.

Integration with telescope environments is achieved through dedicated interface
services. A site-specific control interface translates telescope control system
(TCS) commands into EDD-compatible control messages, allowing the backend to
operate within heterogeneous control infrastructures. In addition, a
site-specific Telescope Metadata Interface (TMI) provides access to
observational metadata -- such as source coordinates or scan parameters -- that
pipelines can consume as required.

The deployment of an EDD backend at a given site is defined by a site
configuration that describes the available hardware resources and the mapping
of services and pipelines onto them. All services and pipelines execute in
isolated runtime environments. This ensures that individual components can be
developed, deployed, and operated independently while maintaining a consistent
execution environment across sites. A schematic overview of these relationships
is shown in Figure~\ref{fig:edd_concept}.

\subsection{Execution and State Model}
\label{sec:state_model}
\begin{figure}
\centering
\includegraphics[width=0.45\textwidth]{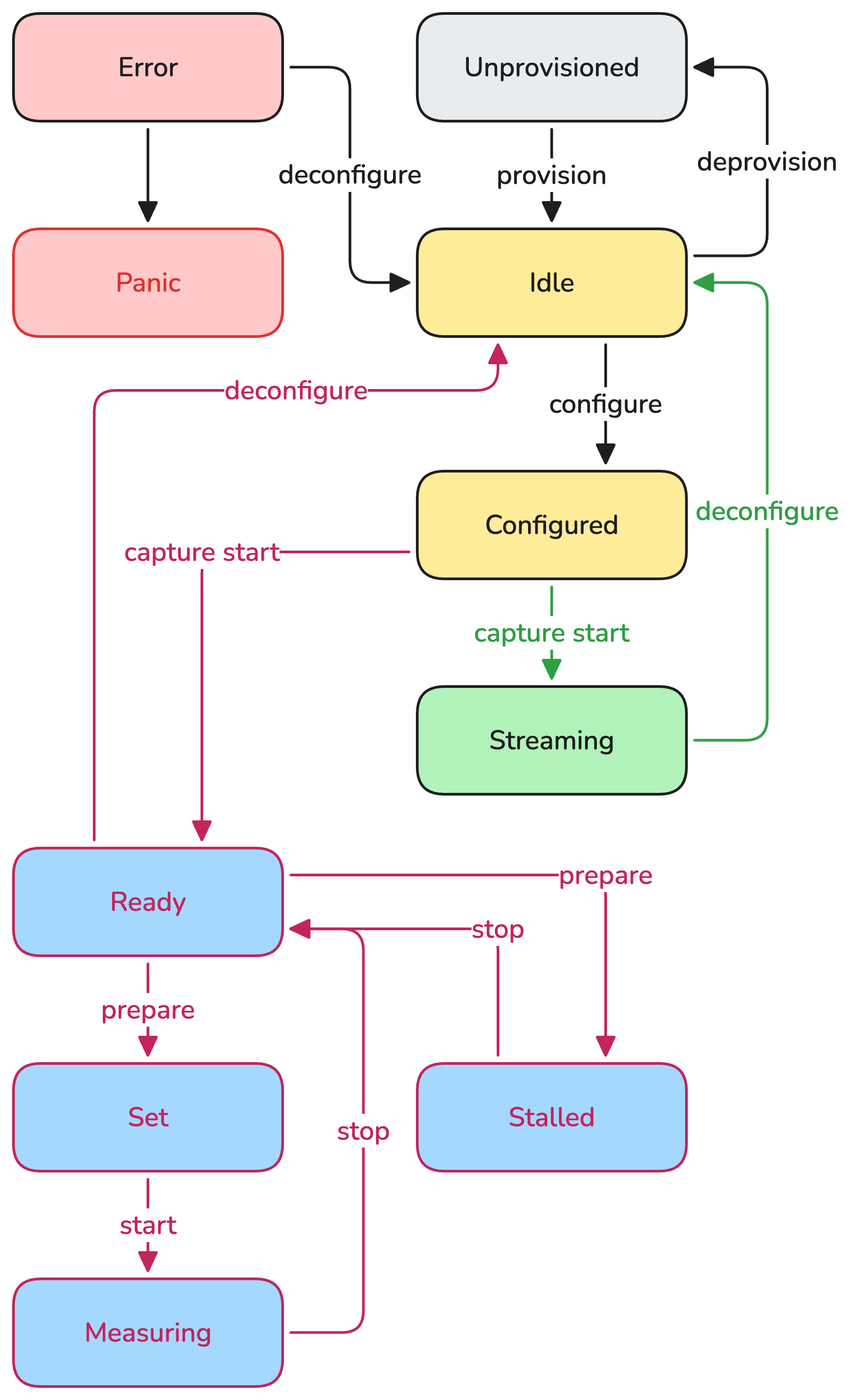}
\caption{State model of the EDD backend. States in which configuration can be
	applied are marked in yellow. Transitional states are not shown for clarity.
	The error state can be reached from every other state and every transition.
\label{fig:edd_state_model}}
\end{figure}

All EDD pipelines adhere to a common state model
(Fig.~\ref{fig:edd_state_model}) that governs their lifecycle and operational
control. This model applies only to pipelines; auxiliary services are managed
independently and do not participate in the observational state machine. The
model is organized as a hierarchy of three control loops, each corresponding to
a distinct reconfiguration timescale in radio astronomical observations.
It distinguishes between parameters defined at pipeline
initialization and those applied on a per-observation or per-scan basis. This
separation allows efficient reconfiguration of pipelines without requiring full
reinitialization, and supports the operational requirements of both continuous
and scan-based observing modes.

At the outermost level, pipelines transition between the \textit{unprovisioned} and
\textit{idle} states. In the \textit{unprovisioned} state, no pipeline
processes are active and the pipeline consumes no resources. The \texttt{provision} command instantiates pipelines on
allocated hardware resources and transitions them to \textit{idle},
establishing the mapping between processing components and infrastructure.
Provisioning reuses the site-specific configuration described in Section \ref{sec:system_architecture}, so that resource
assignments and service descriptions are maintained in a single place. The
reverse transition, \texttt{deprovision}, tears down all pipeline processes and
returns the system to \textit{unprovisioned}.

Starting from the \textit{idle} state, the
configuration loop prepares pipelines for a specific observing mode. The
\texttt{configure} command initializes processes, allocates resources, and
defines input/output data streams, advancing the pipeline to the
\textit{configured} state. Because a pipeline's configuration may modify its
output data streams, configuration is applied by the Master Controller in
topological order of the pipeline graph to ensure that downstream pipelines
receive consistent stream descriptions. From \textit{configured}, the
\texttt{capture\_start} command activates data capture, transitioning the
pipeline into one of two operational states depending on its class.
\textit{Streaming} pipelines operate continuously across multiple scans and
enter the \textit{streaming} state directly. \textit{Measuring} pipelines are
scan-driven and enter the \textit{ready} state, awaiting per-scan parameters.
Both classes return to \textit{idle} via the \texttt{deconfigure} command.

The innermost loop governs per-scan execution for measuring pipelines. The
\texttt{measurement\_prepare} command applies scan-specific parameters -- such
as source coordinates or output targets -- and advances the pipeline to the
\textit{set} state. From \textit{set}, the \texttt{measurement\_start} command
initiates data acquisition, transitioning the pipeline to \textit{measuring}.
On completion, \texttt{measurement\_stop} returns the pipeline to
\textit{ready} for the next scan. These transitions are designed to be
low-latency to support tight observational cadence.

For data provenance, all state transitions are internally indexed, providing each pipeline with a unique identifier for the data products generated during a given measurement. All state transitions, except those into failure or stalled states, are triggered by explicit commands issued by the Master Controller. In normal operation, these commands are generated in response to requests from the telescope control system or from an operator-facing EDD control interface.

The model defines three failure modes: an
\textit{error} state, reachable from any operational state, indicating a fault
expected to be resolved by reiterating the enclosing control loop; a
\textit{stalled} state for cases where a pipeline cannot operate on a
particular scan but is expected to recover on the next; and a \textit{panic}
state, entered when deconfiguration itself fails, indicating that recovery may
require manual intervention.
%The model therefore provides deterministic
%orchestration while allowing different layers of the system to evolve at their
%respective timescales.

\subsection{Communication and Dataflow Model}
\label{sec:comms_and_dataflow}
Communication within the system is separated into control and data channels.
Control communication is based on imperative command messages issued by the
Master Controller to pipelines, defining state transitions and configuration
changes. In addition, pipelines provide asynchronous feedback in the form of
status updates, alerts, and error notifications, allowing the controller to
monitor system health and react to failures.

The dataflow within the EDD backend is realized through the data streams that
connect pipelines in the processing graph. These streams define the interfaces
between pipelines and specify how data is exchanged along the edges of the
graph. Data streams are explicitly described entities that define the contract between
connected pipelines. It is the responsibility of the subscribing party to
ensure it supports the provided data format.

Each pipeline declares its input and output streams to the
Master Controller as part of its configuration. Input streams are referenced by
logical identifiers that are resolved by the controller to compatible upstream
output streams, allowing connections to be established without embedding
deployment-specific details in the pipeline configuration.
Output streams are fully specified by the producing pipeline. Their definition
provides all information required for downstream pipelines to receive and
interpret the data, independent of the underlying transport mechanism. This
includes the structure and semantics of the data stream, as well as any
parameters required for its identification and access. Where applicable, stream
definitions may also describe subdivisions of the data (e.g., by channel, time,
or polarization), enabling downstream pipelines to selectively consume subsets
of a stream.

The EDD architecture is intentionally agnostic with respect to the underlying
data transport mechanisms. Pipelines may exchange data using a variety of
methods depending on the requirements of the deployment environment and the
processing task. Concrete implementations of data transport are discussed in
Section~\ref{sec:data_transport}.

\subsection{
  Extensibility and Customization
}
\label{sec:plugin_model}
The EDD backend is divided into a core framework and an ever-expanding set of plugins. The core provides the essential services described above -- the Master Controller, interface services and monitoring services -- while plugins supply processing pipelines, auxiliary services, and additional components such as data visualization tools or archival interfaces. Plugins are the mechanism by which new observing modes are introduced into the system, and their development and deployment is independent of the core.
An EDD deployment is tailored to the needs of its respective observatory by defining the active plugin set and their versions, allowing different sites to maintain distinct configurations from a common codebase. An overview of currently available plugins and the capabilities they provide is given in Section~\ref{sec:edd_capabilities}.

\section{Implementation and Operations}
\subsection{Deployment and Provisioning of Observation Modes}
\label{sec:deployment_model}
The EDD backend is deployed and maintained using Ansible\footnote{\url{https://docs.ansible.com/}}, an
open-source tool for automation of software provisioning, configuration management, and
application deployment. All site-specific information -- hardware inventories,
network topology, resource assignments, and service configurations -- is
captured in a structured Ansible inventory. Deployment steps are encoded as
Ansible playbooks and role descriptions. On execution, Ansible applies the
roles specified in a playbook to the hosts or host groups defined in the
inventory. From these two sources, which are typically
YAML\footnote{\url{https://yaml.org/spec/1.2.2/}} files, Ansible idempotently
executes all required actions on the target machines so that the complete
installation of
an EDD instance is reproducible. This realises a
configuration-as-code approach~\citep{Morris2025} in which the full system
state is version-controlled, providing both reproducibility and a record of the
deployment history for data provenance.

The EDD core and each plugin are packaged as Ansible collections, the standard
distribution unit for Ansible content. A collection bundles the roles,
playbooks, and supporting files needed to install and operate its components.
The core collection provides the Master Controller, interface services,
monitoring infrastructure, and shared tooling; plugin collections supply
processing pipelines and auxiliary services. The EDD core provides templates
and tooling to assist in developing plugin collections.

%This organisation
%mirrors the core/plugin separation described in Section~\ref{sec:plugin_model}:
%the active set of collections and their pinned versions fully defines a site's
%EDD installation.

During installation container images are pulled from a central registry,
configuration files are generated from the inventory, and supporting
infrastructure such as databases and monitoring services is started. All EDD
services run in individual Docker\footnote{\url{https://www.docker.com/}} containers, providing isolated
environments in which each service's dependencies are bundled independently.
This OS-level virtualisation decouples services from one another and from the
host operating system, allowing an EDD installation to integrate into diverse
computing environments without imposing constraints on the host distribution or
library versions.
For sites that manage their configuration in a git repository, the core
provides a tool to deploy tagged versions from the repository via a single
command from a clean environment. The \emph{edd-tool} handles all required git checkouts
and calls to Ansible. The tool is installable from the python packaging index.

%\subsection{Provisioning of Observation Modes}
After the EDD is deployed at a site, individual observation modes can be provisioned by the user using commands sent to the Master Controller.
When an observation is initiated, the Master Controller executes Ansible
playbooks selected for the requested observing mode, using the same
site-specific inventory as in the installation phase. These playbooks launch
the required pipeline and service containers on their designated resources,
establishing the mapping between processing components and hardware. This reuse
of a single inventory ensures that resource assignments remain consistent
between installation and runtime provisioning, and avoids maintaining parallel configuration
sources. Deprovisioning reverses this process, tearing down
observation-specific containers and returning the system to its initial state.

The distinction between installation and provisioning reflects a separation of
timescales: installation is performed infrequently by engineering staff and may
involve infrastructure changes, whereas provisioning and deprovisioning occur
routinely as part of the observational cycle and are designed to complete
within the timescales required by telescope operations, typically a few minutes.

\subsection{
  Control and Protocols
}
\label{sec:control_and_protocols}
Communication between the Master Controller and all pipelines uses the KAroo
Telescope Communication Protocol (KATCP)\footnote{\url{https://pythonhosted.org/katcp/}}, a text-based protocol
developed by SKA South Africa for control and monitoring of devices in the
MeerKAT telescope. In KATCP, devices and controllers exchange newline-separated
messages over a TCP/IP stream, providing a lightweight mechanism for issuing
commands, querying device state, and receiving asynchronous status updates.
Unlike stateless REST-over-HTTP interfaces, KATCP maintains a persistent
connection over which devices can push sensor updates without being polled,
avoiding the need for auxiliary feedback channels. The choice of KATCP was
motivated by prior experience with the protocol during development of the
APSUSE and FBFUSE instruments for MeerKAT~\citep{fbfuse, fbfuse2} and the
availability of mature third-party libraries, in particular
\texttt{aiokatcp}\footnote{\url{https://github.com/ska-sa/aiokatcp}}, which provides a fully asynchronous Python
implementation suitable for integration into the EDD's concurrent control
architecture.

Within the EDD, all pipelines expose a common set of KATCP commands
corresponding to the state transitions defined by the state model
(Section~\ref{sec:state_model}), as well as a standardised set of sensors
reporting pipeline status, performance metrics, and diagnostic data. Pipelines
may extend this interface with additional mode-specific commands and sensors.
For Python-based pipelines, the EDD core provides an abstract base class,
\texttt{EDDPipeline}, that implements the common command and sensor interface;
pipeline developers extend this class to provide their processing logic.
Pipelines implemented in other languages need only conform to the KATCP wire
protocol.

Infrastructure-level operations -- container deployment, host configuration, and
service installation -- are managed by Ansible over Secure Shell (SSH), as described in
Section~\ref{sec:deployment_model}. KATCP and SSH thus operate at distinct
levels of the system: SSH provides the transport layer for deployment and
provisioning actions executed by Ansible, while KATCP provides the runtime
control and monitoring layer through which the Master Controller orchestrates
pipeline behaviour during observations.

Integration with external telescope control systems that cannot connect directly to the Master Controller is handled by dedicated,
site-specific interface services as introduced in
Section~\ref{sec:system_architecture}. For command-driven control, this interface translates commands from the local telescope control system to KATCP messages for the Master Controller. At Effelsberg,
for example, a service accepts SCPI\footnote{\url{https://www.ivifoundation.org/downloads/SCPI/scpi-99.pdf}} commands from the existing
telescope control infrastructure. For metadata access, a TMI service translates
site-specific data sources into a format consumable by EDD pipelines; at
Effelsberg, this service subscribes to the local metadata stream via
ZeroMQ\footnote{\url{https://zeromq.org/}}. Both interface types are intentionally thin translation
layers, imposing no constraints on the upstream protocol beyond the requirement
that a mapping to and from KATCP can be implemented. This design results in
minimal development overhead when integrating EDD backend instances at
observatory sites.

\subsection{Data Transport}
\label{sec:data_transport}
As described in Section~\ref{sec:comms_and_dataflow}, the EDD framework is
intentionally agnostic to the data transport mechanism used between pipelines.
Pipelines declare well-defined input and output stream descriptors during the
configuration phase, and any transport that satisfies these descriptors can be
used. However, in practice, the active processing capabilities within the EDD
ecosystem have converged on a common set of protocols and tooling, which we
describe here. The primary data transport method is IP multicast~\citep{deering1989ipmulticast} over
Ethernet. In this model, pipelines publish data products to one or more
multicast groups to which downstream pipelines subscribe. The subscription
mechanism is handled at the network layer: datagrams are replicated by the
switching infrastructure only when multiple subscribers are present, making
fan-out efficient without requiring the producing pipeline to be aware of the
number or identity of its consumers. This publish-subscribe model decouples
producers from consumers and simplifies the construction of complex processing
graphs in which a single data stream may feed multiple independent downstream
stages. The EDD core provides tooling for working with multicast traffic,
including automatic management of multicast group address allocation to avoid
conflicts within a deployment. The payload format used over multicast in most
EDD deployments is the Streaming Protocol for the Exchange of Astronomical Data
(SPEAD)\footnote{\url{https://casper.berkeley.edu/astrobaki/images/9/93/SPEADsignedRelease.pdf}}. SPEAD encodes self-describing items -- comprising both
metadata and bulk data -- into UDP packet streams, supporting complex
multidimensional data types and flexible many-to-many stream topologies. Its
self-describing nature allows consumers to interpret incoming data without
out-of-band schema negotiation, which is well suited to the dynamic pipeline
topologies that arise when the EDD is reconfigured between observing modes.
High-performance transmission and capture of SPEAD streams is provided by the
tools \texttt{mksend} and \texttt{mkrecv}, developed in parallel with but
independently of the EDD core for the FBFUSE, APSUSE, and TUSE instruments at
MeerKAT. These tools are built on the \texttt{spead2} library\footnote{\url{https://spead2.readthedocs.io/en/latest/}},
which provides an optimised C++ implementation of the SPEAD protocol with
Python bindings. More recent versions are being migrated to use the Data Plane
Development Kit (DPDK)\footnote{\url{https://www.dpdk.org/}} for low-level packet handling, bypassing the
kernel networking stack to achieve the line-rate capture and transmission
performance required at high data rates. Both tools are currently being
consolidated into a unified application, \texttt{eddio}, which will provide a
single interface for high-performance stream I/O across EDD deployments. Using
these tools we achieve 95\% line-rate transmission and reception when testing over
200-GbE connections. For VLBI applications, the EDD additionally supports
transmission and recording using the VLBI Data Interchange Format
~\citep[VDIF,][]{whitney2010vdif}, as described in Section~\ref{sec:vlbi}.

\subsection{Dependency and Change Management}
As with any modern software project, the EDD depends on numerous external
open-source libraries, each with its own dependency tree. Managing these
dependencies is critical to the continuous functionality of the system. Three
concerns must be addressed: compatibility of all library versions across the
software stack, particularly where libraries share common transitive
dependencies; continued availability of dependencies in their required
versions, since upstream projects may be discontinued, their repositories
deleted, or their software modified without correct version tracking; and
licence compatibility across all components.
The EDD addresses these concerns through three complementary measures:
organisation in plugins, containerisation of services, and packaging of
software components.

As described in Section~\ref{sec:deployment_model}, each plugin's code runs in
dedicated containers that communicate only through well-defined interfaces
(Sections~\ref{sec:control_and_protocols} and~\ref{sec:data_transport}). This
achieves a high degree of code decoupling, though it requires that these
interfaces remain stable. To date, no backwards-incompatible changes to the
communication interfaces have been necessary since their introduction.
Within this boundary, the internal dependencies of plugins and the core must
still be managed coherently. Linux distributions address precisely this problem
by bundling software versions that are tested to work without conflicts.
For the EDD core and all plugins developed by the principal EDD development team,
we build all software against fixed suites -- defined combinations of a Linux
distribution and CUDA version, with dependencies installed from the
distribution's repositories. Where required packages are not available
upstream, we create Debian\footnote{\url{https://www.debian.org/}} packages and
distribute them via a dedicated
repository\footnote{\url{https://debinstall.mpifr-bonn.mpg.de/edd-deb/}}. At the time
of writing, the default suite is Ubuntu 24.04 with CUDA 12.6.1; we intend to
follow the Ubuntu Long Term Support release cycle. From these packages,
container images for all services are built and published to a central
registry\footnote{\url{https://gitlab.mpcdf.mpg.de/groups/mpifr-bdg/-/container_registries}}
via an automated CI/CD pipeline.
Using binary software packages here also makes the deployment independent on
the availability of upstream resources and upstream change-management compared
to individual on-site builds. Furthermore, it reduces the size of container
images as build dependencies and source trees are not required. This reduces
the time needed to switch between observation modes.

The automatic CI/CD pipelines enables frequent version updates.
The EDD relies on hardware
accelerators such as GPUs whose driver and CUDA support mandates continuous
updates; a static dependency stack would quickly become incompatible with
current hardware and unable to benefit from improvements in upstream libraries.
The system is therefore designed to integrate frequent small changes
continuously, rather than accumulate infrequent large updates that carry
greater risk of breakage.
Sustaining this pace of change requires comprehensive automated testing. Tests
operate at three levels: unit tests within individual packages and plugins;
integration tests that exercise individual pipelines through the full state
model command sequence; and system-level tests run against complete EDD
deployments, in which a test manager loads provision descriptions and executes
a full measurement cycle. After each state transition, pipeline-specific checks
verify correct system behaviour. A report recording all results is generated
automatically, including diagnostic plots that are reviewed by a human operator
to catch issues not amenable to automated verification.

\subsection{Monitoring and Logging}
For operating the EDD it is important to capture the state of individual processing pipelines as well as their performance parameters, both to ensure operational performance and to record metadata needed to assess the quality of recorded data.
Every pipeline exposes a set of default metrics -- such as pipeline status, processing throughput, and buffer occupancies -- as well as pipeline-specific data using KATCP sensors. The sensor data is collected from pipelines by sidecars: additional monitoring services that are automatically deployed alongside every pipeline during provisioning. The collected data is injected into two databases serving complementary roles. An in-memory database provides low-latency access to the current system state and can store large data objects such as diagnostic plots exposed by pipelines. A time-series database records high-resolution scalar metrics -- such as network packet loss rates or processing latencies -- that are collected within the sidecar and written to the database in batches. The separation reflects differing requirements: the online view requires rapid access to potentially large objects at modest time resolution, while retrospective analysis and data quality assessment require high-resolution time series of key performance indicators.
The aggregated data from both systems is used to provide a real-time view of the system for human and automated operators. As an example, a spectrometer pipeline may expose its current bandpass as a diagnostic plot in the in-memory database for immediate visual inspection, while the corresponding packet loss statistics are recorded at high resolution in the time-series database for later use in estimating data quality.
As in-memory database the EDD uses Redis\footnote{\url{https://redis.io/}}; as time-series database, InfluxDB version 1.8\footnote{\url{https://www.influxdata.com/}}. Data from both systems is visualised using Grafana\footnote{\url{https://grafana.com/}}. In addition to default dashboards that display the overall state of the system, dedicated dashboards for individual pipelines are dynamically injected into the Grafana instance to adapt the monitoring view to the current observing mode. All pipelines log to standard output. Logs are collected from the Docker daemon and ingested into a Loki\footnote{\url{https://grafana.com/oss/loki/}} instance for centralised querying and analysis. Where required, logs can additionally be forwarded to a site-specific log aggregator such as Graylog\footnote{\url{https://graylog.org/}}. This approach -- centralised collection from container standard output rather than application-level log management -- ensures a uniform logging model across all pipelines and services regardless of their implementation language or internal structure.

\section{Observing Modes}
\label{sec:edd_capabilities}

The EDD backend currently supports a broad range of observing modes through its plugin architecture. Available capabilities include digitisation and packetisation for several digital receiver systems, channelisation, spectroscopy and spectropolarimetry, pulsar timing, pulsar and transient searching, baseband recording, Very Long Baseline Interferometry (VLBI), correlation and beamforming, and various output formatting pipelines for formats including HDF5\footnote{\url{https://www.hdfgroup.org/solutions/hdf5/}}, PSRFITS~\citep{psrfits}, and VDIF~\citep{whitney2010vdif}. Each capability is implemented as one or more plugins that supply the processing pipelines and auxiliary services required for the corresponding mode. In the following, we describe three representative modes in detail.

\subsection{Spectroscopy and Spectropolarimetry}
\label{sec}

\begin{figure}[htbp]
\centering
\includegraphics[width=0.5\textwidth]{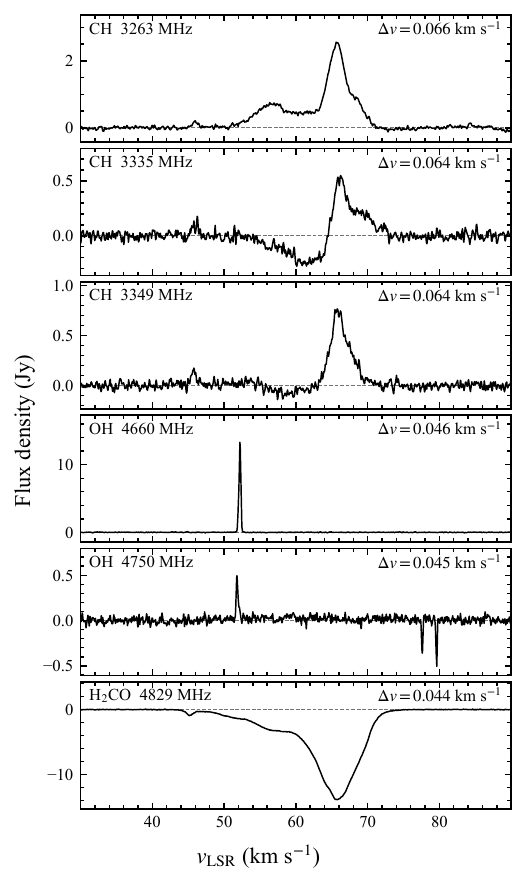}
\caption{Simultaneous spectroscopic zoom-mode observations of CH, OH, and H$_2$CO transitions in the W51 star-forming region with the Ultra Broadband (UBB) receiver on the Effelsberg radio telescope. The spectra are plotted against the Local Standard of Rest (LSR) velocity, with the velocity resolution ($\Delta v$) indicated.}
\label{fig}
\end{figure}

The spectroscopy pipeline provides frequency- and polarisation-resolved measurements for continuum, spectral-line, and polarimetric observations. It implements GPU-accelerated channelisation and cross-polarisation coherence detection, enabling the simultaneous recovery of intensity and polarisation products across the full bandwidth of modern wideband receivers. The pipeline employs a gated architecture in which the incoming data stream is partitioned according to the state of the receiver noise diode, as encoded in the digitised data stream. This allows for simultaneous production of science data along with real-time calibration, without requiring offline reconstruction of the switching cycle. The approach supports rapid noise-diode switching at rates up to the resolution of the gate flag, typically on microsecond timescales, and is therefore well suited to continuum and polarimetric observations requiring time-resolved gain calibration. It also enables commensal operation with time-domain modes such as pulsar searching, since the switching period can be chosen to be shorter than the time resolution of the commensal data products, avoiding periodic artefacts. While the noise-diode switching mode is principally designed for EDD receivers with inbuilt noise diodes driven via the digitiser, it is also usable with an external noise source provided that the EDD backend has access to the generated waveform.

Channelisation is performed on GPUs using an FFT-based filterbank. For a single dual-polarisation \SI{3}{\giga\hertz}-wide input band, the system provides up to 32 million spectral channels in intensity mode, corresponding to a frequency resolution of approximately \SI{100}{\hertz}. At L-band (\SI{\sim 1.4}{\giga\hertz}), this corresponds to a velocity resolution of approximately \SI{0.02}{\km\per\second}, suitable for high-resolution spectral-line studies. When full cross-polarisation coherences are computed for Stokes parameter determination, up to 8 million channels are supported at the same bandwidth, corresponding to a velocity resolution of approximately \SI{0.08}{\km\per\second} at L-band. These configurations are achieved using at most four GPUs with current consumer-grade (NVIDIA RTX~3090) or data-centre (NVIDIA L40) hardware. Scaling to larger aggregate bandwidths can be achieved either by increasing per-node processing and network ingest capacity, or by distributing the workload across multiple pipeline instances operating on disjoint frequency subbands.

The pipeline can also be operated in a spectroscopic zoom mode, in which high spectral resolution is retained only around selected frequency ranges. This mode is useful for wideband receivers where the instantaneous bandwidth contains multiple transitions of interest, but where recording the full band at the highest spectral resolution would produce unnecessary data volume. As an example, Fig.~\ref{fig} shows observations of the W51 star-forming region obtained with the upper band of the Effelsberg Ultra Broadband (UBB) receiver, covering \SIrange{3}{6}{\giga\hertz}. The observations targeted several CH, OH, and H$_2$CO transitions simultaneously. In this configuration, the spectrometer used $2^{22}$ channels across the input band, giving a channel spacing of approximately \SI{715}{\hertz}. The retained zoom windows were approximately \SI{70}{\mega\hertz} wide around each targeted transition, corresponding to velocity resolutions between \SI{0.044}{\km\per\second} and \SI{0.066}{\km\per\second} over the observed frequency range. The observations were performed in position-switching mode. During reduction, corrections for atmospheric attenuation were applied, and the flux-density scale was established by comparison with continuum observations of standard calibrators such as 3C48 and NGC~7027.

Spectrometer output data can be recorded directly to disk in raw binary format or forwarded to dedicated downstream pipelines. Separate plugins provide recording in HDF5 format and forwarding to an external APEX-compatible FITS writer~\citep{Muders2018}, the latter enabling integration with existing observatory data flows at Effelsberg. The spectroscopy plugin is deployed operationally at Effelsberg, the SKA-MPI prototype and demonstrator dish (SKAMPI), and the Thai National Radio Telescope (TNRT). At Effelsberg, all receiver configurations include a spectrometer instance feeding the telescope's FITS writer, providing a uniform spectroscopic capability across the observatory's receiver suite. At TNRT, the spectrometer was the first commissioned observing mode and has been used to detect OH maser emission from a comet~\citep{Sakai2025comet}. At SKAMPI, the spectrometer forms the basis of an ongoing total-power survey of the southern sky in S-band~\citep{Juenemann2026}.

\subsection{Pulsar Timing and Searching}
\label{sec:pulsar}

\begin{figure*}
\centering
\includegraphics[width=\textwidth]{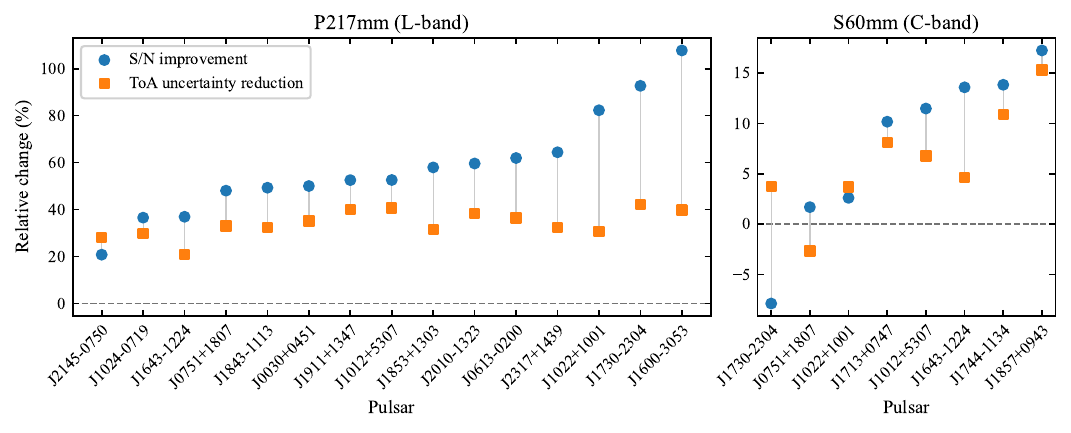}
\caption{Relative change in signal-to-noise ratio (S/N; circles) and pulse time-of-arrival (ToA) uncertainty (squares) for simultaneous EDD and PSRIX observations of well-characterised European Pulsar Timing Array pulsars. Positive values indicate improved S/N or reduced ToA uncertainty for EDD relative to PSRIX. The left and right panels show the P217mm and S60mm receivers respectively, which differ in whether EDD digitises at the telescope focus or in the Faraday room (Sect.~\ref{sec:pulsar}). Where multiple simultaneous epochs are available for a given pulsar, the mean is shown.}
\label{fig:pulsar}
\end{figure*}

Pulsar observations typically require coherent dedispersion over wide bandwidths, precise time tagging referenced to observatory clock standards, and support for both real-time analysis and high-throughput recording. The EDD pulsar plugin provides timing, search, and baseband recording modes within a single configurable framework. It supports two complementary ingest paths, selected according to the incoming data rate. For moderate bandwidths, unchannelised dual-polarisation voltage streams are ingested directly by the processing pipeline. For higher data rates exceeding the capacity of a single processing node, an upstream channelisation stage divides the signal into frequency subbands, which are distributed across multiple processing nodes using the multicast data transport described in Section~\ref{sec:data_transport}. Although this channelisation step is implementation-independent in principle, the EDD provides a CASPER-based FPGA polyphase filterbank plugin for this purpose, illustrating that the plugin architecture can incorporate heterogeneous hardware components as well as software pipelines. Both ingest paths produce data in a common internal format and are handled transparently by the downstream processing stages.

In \textit{timing} mode, the pipeline performs GPU-accelerated coherent dedispersion and phase-aligned folding using \texttt{dspsr}~\citep{dspsr} and \texttt{tempo2}~\citep{tempo2}. When a valid pulsar ephemeris is available, the folding predictor is generated automatically; the ephemeris repository can be specified at runtime, allowing observers to select project-specific or curated timing solutions. In \textit{search} mode, the pipeline produces coherently dedispersed filterbank data using \texttt{digifits}~\citep{dspsr}, with configurable time and frequency resolution. RFI mitigation, including spectral channel excision, is applied as part of the processing chain, and output is written in standard PSRFITS~\citep{psrfits} format. In \textit{baseband} mode, the raw complex voltage stream is recorded directly to disk as DADA files\footnote{\url{https://dspsr.sourceforge.net/manuals/dspsr/dada.shtml}}, preserving the full signal information for offline reprocessing. This enables analysis techniques such as voltage-domain RFI excision and coherent-dedispersion-enabled fast transient searches \citep[e.g.][]{BASSA201740}.

All modes share a common configuration interface and expose a consistent set of monitoring sensors. Data products are organised automatically using observation metadata, and post-processing steps, including profile scrunching, channel excision, and diagnostic plotting, are triggered automatically as data products are created. On current data-centre hardware (NVIDIA L40 GPUs), the pipeline sustains coherent dedispersion and folding at \SI{\sim1}{\giga\hertz} of bandwidth per GPU.

Pulsar timing and searching capabilities have been available at Effelsberg since 2019. Observations using this mode have contributed to studies of scintillation \citep{main2023scintillation,main2023frbscintillation}, scattering screen dynamics \citep{sprenger2026scintillometry}, FRB emission and progenitor environments \citep{eppel2025frb20240114a,nimmo2025magnetosphere,braga2025frb20121102a,pearlman2025frb,limaye2026ubbfrb}, and magnetar emission properties \citep{younes2025magnetar}. The next data release for the European Pulsar Timing Array \citep[EPTA,][]{Jawor2026epta} will contain EDD pulsar timing measurements.

We assessed the comparative sensitivity of EDD and the legacy PSRIX backend~\citep{lazarus2016psrix} using simultaneous observations of well-characterised millisecond pulsars from the EPTA programme with two Effelsberg receivers. In both cases a common processing bandwidth and RFI mitigation strategy was used between backends. The P217mm L-band, \qtyrange{1510}{1250}{\mega\hertz}, receiver provides both analogue and digital outputs. For this comparison, the EDD signal was digitised at the telescope focus and transported digitally to the backend over Ethernet, while the analogue output was transmitted through approximately \SI{400}{\metre} of RF coaxial cable and digitised by PSRIX in the observatory Faraday room. As a control, we performed an analogous comparison with the S60mm C-band receiver, \qtyrange{4600}{5100}{\mega\hertz}, which has no in-receiver digitiser: here an EDD digitiser was installed in the Faraday room and supplied with the same analogue input as PSRIX, so that both backends operate on identical signals and the comparison isolates backend-intrinsic differences.

The results are shown in Fig.~\ref{fig:pulsar}. For the P217mm receiver, where the comparison combines the effect of digitising at the telescope focus with any backend-intrinsic differences, EDD shows a clear improvement over PSRIX: the measured signal-to-noise ratios increase by 20--110\%, and the corresponding pulse time-of-arrival (ToA) uncertainties are reduced by approximately 20--40\%. The fractional ToA improvements are smaller than the radiometer-noise scaling of the S/N gains would imply, and depend only weakly on the magnitude of the S/N gain. This is consistent with the timing precision of these bright millisecond pulsars being partly limited by pulse-phase jitter and finite-template effects rather than by radiometer noise alone, so that gains in raw sensitivity are not fully realised as reductions in ToA uncertainty.

For the S60mm receiver, where both systems digitise in the Faraday room after identical analogue transmission, the differences are much smaller. EDD shows a modest, predominantly positive tendency, with S/N changes ranging from about $-8\%$ to $+17\%$ and comparably small changes in ToA uncertainty. We therefore attribute the bulk of the P217mm improvement to digitising close to the receiver, which avoids signal degradation over the approximately \SI{400}{\metre} analogue transmission path. The S60mm comparison indicates that backend-intrinsic differences between EDD and PSRIX are small relative to this effect, although it does not establish exact parity: a residual tendency at the ten per cent level in favour of EDD remains, and the two comparisons are not matched in observing band, so the C-band result is an imperfect control for the L-band case. A detailed characterisation of this residual difference is beyond the scope of this work.

\subsection{Very Long Baseline Interferometry}
\label{sec:vlbi}

% \begin{figure*}
% \centering
% \includegraphics[width=\textwidth]{figures/VLBI_fringe_EFF_TNRT.pdf}
% \caption{First successful VLBI fringe detection with the TNRT, correlated using the DiFX software correlator. The fringe plot shows the correlated visibility amplitude and phase as functions of delay and fringe rate, demonstrating a clear fringe detection between TNRT and Effelsberg. This result marks the first confirmed VLBI detection for TNRT and demonstrates the successful integration of the EDD's VLBI mode \citep{KSugiyama2024EVN}.\textcolor{red}{This is not the final figures.}\label{fig:vlbi_fringe_eff_tnrt}}
% \end{figure*}

VLBI observations require a backend to produce precisely timed, standardised data streams that are compatible with the recording and correlation infrastructure of international VLBI networks. The EDD VLBI plugin provides this capability by converting digitised wideband signals into multiple VLBI-compatible sidebands, packaging them in standard VDIF format, and either recording them locally or streaming them to external recorders.
The core of the VLBI capability is a GPU-accelerated digital down-converter (DDC) pipeline that extracts arbitrary sidebands from the wideband input signal. For each requested sideband, the input is mixed with a configurable local-oscillator frequency and resampled to the target output rate using a polyphase resampler, which allows efficient extraction even when the target rate is not an integer divisor of the input sampling frequency. Upper and lower sidebands are generated via a Hilbert transform applied to the resulting analytic signal. A separate VDIF packing pipeline accepts the DDC output -- or, where no down-conversion is required, raw packetised data directly -- re-quantises to 2- or 8-bit precision, and prepends VDIF headers. The packer can record to local disk or transmit to external recording systems. For demanding configurations, such as multi-band or multi-feed receivers, the packing stage can also be executed on a GPU.
Additionally, the VDIF packaging pipeline provides live spectral waterfall displays in the Grafana monitoring interface.

The EDD can also include additional pipelines alongside the DDC and VDIF processing. In particular, spectroscopic modes can be deployed commensally with the VLBI pipeline, allowing spectral and continuum products to be produced without interrupting the VLBI backend configuration. This is useful for observing schedules in which short spectroscopic or continuum measurements are required between VLBI scans, for example to determine system temperature, observe calibrators, or support antenna pointing and focus checks. Provided that sufficient compute and network resources are available, these capabilities are provisioned together and enabled or disabled as required by the observing sequence; no backend reprovisioning cycle is required between the VLBI and spectrometer uses.

Integration with the VLBI Field System\footnote{\url{https://github.com/nvi-inc/fs}}, the standard control software for VLBI stations, is provided by a dedicated interface service that translates Field System commands into KATCP messages for the EDD Master Controller, emulating both a DBBC backend and a Mark recording system from the Field System's perspective.
The VLBI plugin is deployed at both Effelsberg and SKAMPI. At SKAMPI, it has been used for successful VLBI observations with international networks~\citep{Jompoj2026}, demonstrating that the EDD can serve as a complete VLBI backend for stations that lack legacy VLBI-specific hardware. At Effelsberg, the VLBI capability operates alongside the observatory's existing DBBC systems, providing a software-defined alternative that can be reconfigured for non-standard sideband selections or bandwidths without hardware changes.

\section{EDD Deployments}

The EDD backend is currently deployed at four sites: the Effelsberg 100-m radio
telescope, a large single-dish observatory with extensive legacy
infrastructure; SKAMPI, a small
robotic single-dish operating with minimal human supervision; the TNRT, a 40-m dish in Chiang Mai and the largest radio
telescope in South-East Asia, with frequency coverage from \SIrange{0.3}{115}{\giga\hertz} and
the 5-antenna ARGOS interferometric prototype array, currently in
commissioning. These deployments span a wide range of scales, network
architectures, and digitisation technologies, while operating a common EDD
codebase differentiated only by site-specific configuration and plugin
selection. Table~\ref{tab:deployments} summarises the configuration at each
site. As the TNRT deployment is similar in scale and design to SKAMPI, we focus
in the following on the remaining three sites.

\begin{table*}
\centering
\caption{Hardware and infrastructure summary of EDD deployments. NDR IB denotes NVIDIA NDR InfiniBand.\label{tab:deployments}}
\small
\begin{tabularx}{\linewidth}{l X X X}
\hline
 & \rule{0pt}{2.6ex} \textbf{Effelsberg (EDGAR)} & \textbf{SKAMPI} & \textbf{ARGOS Prototype}\rule[-1.0ex]{0pt}{0pt} \\
\hline
\textbf{Telescope / array}
  & \rule{0pt}{2.6ex}Single 100-m dish
  & Single 15-m dish
  & $5\times$ 6-m dish array \\[6pt]
  \addlinespace[8pt]
\textbf{Operational since}
  & 2024 (predecessor hardware since 2019)
  & 2020
  & in commissioning \\[6pt]
  \addlinespace[8pt]
\textbf{Topology}
  & Disaggregated: dedicated VM hosts, compute nodes, and storage nodes on separate data and storage fabrics
  & Converged: combined processing and storage services but dedicated fabrics for data acquisition and storage
  & Hyper-converged: compute, storage, and networking co-located across three nodes on a shared fabric \\[6pt]
  \addlinespace[8pt]
\textbf{Data ingest and network}
  & $>180\times$ 100~GbE links from primary and secondary focus into aggregators with uplinks to 400/800-GbE spine
  & $4\times2$ 100~GbE from receiver digitisers into 100~GbE with uplink to 100~GbE switch
  & $4\times2$ RFSoC boards, $5\times$ 100~GbE to switch \\[6pt]
  \addlinespace[8pt]
\textbf{Processing hardware}
  & 36 GPU-accelerated compute nodes, 1 FPGA node, 8 VM hosts
	& 2 GPU-accelerated servers, 1 hosting also a FPGA card
  & 3 GPU-accelerated servers \\[6pt]
  \addlinespace[8pt]
\textbf{Storage configuration}
  & $6\times$ nodes ($60\times$ 18\,TB HDD each), BeeGFS on NDR InfiniBand SAN, 6.5\,PB total
  & $24\times$ 4\,TB NVMe each, BeeGFS on FDR InfiniBand SAN, 160\,TB total
  & $8\times$ 16\,TB HDD each, CephFS on 100~GbE, 320\,TB total\\[6pt]
  \addlinespace[8pt]
\textbf{Secondary use}
  & Idle nodes reassigned to HTCondor compute pool
  & Robotic control system and data processing hosted on same hardware
  & Imaging and calibration system hosted on same hardware \\[6pt]
  \addlinespace[8pt]
\textbf{Active observing modes}
  & Spectroscopy, pulsar timing, search and baseband, VLBI
  & Spectroscopy, pulsar timing, VLBI
  & Beamforming, correlation, transient search, pulsar timing and search, VLBI \\
\hline
\end{tabularx}
\end{table*}

\subsection{EDGAR: The EDD at Effelsberg}
The Effelsberg 100-m radio telescope operates a diverse receiver suite
currently comprising 15 systems covering frequencies from 400\,MHz to 95\,GHz,
with instantaneous bandwidths of up to 5\,GHz per polarisation. The suite is
undergoing a transition toward wide-bandwidth digital receivers based on the
EDD frontend system, which performs signal conditioning, down-conversion, and
digitisation in the receiver housing itself. The digitised signals are
packetised into multicast UDP data streams, which are directly ingested by EDD
processing pipelines. Legacy analogue receivers are supported by a digitisation
unit located next to the computing cluster, allowing the full receiver suite to
be served by a unified
backend infrastructure. The most demanding instrument is a 256-element
cryogenically cooled phased array feed (CryoPAF), which is expected to produce
an aggregate data rate of 6.7\,Tbit/s.

To support this diverse receiver suite within a unified backend framework, we
deployed the Effelsberg Data Gathering and Analysis Resource (EDGAR), a
heterogeneous computing cluster housed in the observatory's Faraday room -- a
temperature-controlled, shielded environment that serves as the data centre for
the telescope's processing equipment. The cluster follows a disaggregated
architecture with dedicated tiers for management, GPU and FPGA processing, and
storage, interconnected by separate data and storage network fabrics
(Table~\ref{tab:hardware}). Data from the telescope's primary and secondary
foci is delivered to the cluster over 100-GbE links via an 800-GbE data
acquisition switch.

The network is segmented into isolated virtual networks (VLANs). This is
operationally important for Effelsberg as EDGAR is used not only for
observations but also hardware and software development, and accidental data
streams from the development environment could disrupt production observations.
Production and development nodes are assigned to separate VLANs, so that
erroneous traffic in development is confined and cannot reach production.

The separation of production and development environments is further aided by
hardware virtualisation using a KVM-based hypervisor for the entire cluster.
The hypervisor uses PCI passthrough to provide direct hardware access to
virtual machines for features such as RDMA and GPU compute. Each virtual
machine, together with the network interfaces and accelerators passed through
to it, is assigned either to the production or to the development environment
preventing resource conflicts on the node level. Additional measures for
quality-of-service assurance can be taken on the switch level to prevent for
example the saturation of shard inter-switch connections if necessary. The
hypervisor layer introduces no measurable performance overhead, while providing
benefits in terms of resource flexibility and maintainability.

While EDGAR is designed to accommodate the full CryoPAF data rate, it is
substantially overprovisioned for the data rates of the telescope's
conventional receivers. To utilise this idle capacity, compute nodes not
assigned to an active EDD observation are dynamically reassigned to an
HTCondor\footnote{
  \url {
  https: // htcondor.org/}} high-throughput computing pool running alongside the
EDD. Nodes are added to or removed from the HTCondor pool depending on the
observation schedule, allowing EDGAR to serve as a shared resource for both
real-time telescope operations and general-purpose scientific computing.

Configuration and lifecycle management of the
cluster infrastructure is implemented using Foreman\footnote{\url{https://theforeman.org/}}
and Puppet\footnote{\url{https://www.puppet.com/}}, with Ansible managing the EDD deployment and runtime
provisioning as described in Section~\ref{sec:deployment_model}.
Integration of the EDD into the Effelsberg control environment is achieved
through site-specific interface services
(Section~\ref{sec:control_and_protocols}), namely an SCPI-based command
interface and a ZeroMQ-based TMI server that provides access to telescope
metadata.

\begin{table*}
\centering
\caption{Hardware specification of EDD computing clusters at all deployment sites.\label{tab:hardware}}
\small
\renewcommand{\arraystretch}{1.3}
\begin{tabularx}{\linewidth}{l c l p{1cm} p{2.5cm} X X}
\hline
\textbf{Node type} &
\textbf{Count} &
\textbf{CPU} &
\textbf{RAM} &
\textbf{Local storage} &
\textbf{Network} &
\textbf{Accelerators / other} \\
\hline
\addlinespace[4pt]
\multicolumn{7}{l}{\textbf{EDGAR (Effelsberg 100-m telescope)}} \\
\addlinespace[4pt]
VM hosts &
8 &
$2\times$ Intel Xeon Gold 5318Y &
2\,TB &
16\,TB NVMe &
$3\times$ 400\,GbE \newline $1\times$ NDR IB &
-- \\
\addlinespace[6pt]
GPU processing nodes &
36 &
$2\times$ AMD EPYC 9334 &
768\,GB &
4\,TB SSD &
$2\times$ dual 400\,GbE\newline $1\times$ NDR IB &
$2\times$ NVIDIA L40 \\
\addlinespace[6pt]
FPGA processing node &
1 &
$2\times$ AMD EPYC 9334 &
768\,GB &
4\,TB SSD &
$1\times$ dual 400\,GbE\newline $1\times$ NDR IB &
$8\times$ Xilinx\newline Alveo U55C \\
\addlinespace[6pt]
Storage nodes &
6 &
$2\times$ Intel Xeon Gold 6338 &
512\,GB &
4\,TB NVMe\newline $60\times$ 18\,TB HDD &
$1\times$ dual 400\,GbE\newline $1\times$ NDR IB &
$2\times$ HBA \\
\hline
\addlinespace[4pt]
\multicolumn{7}{l}{\textbf{SKAMPI (SKA-MPI prototype dish)}} \\
\addlinespace[4pt]
Processing nodes &
2 &
$2\times$ Intel Xeon Silver 4516Y &
512\,GB &
$24\times$ 4\,TB NVMe &
$2\times$ dual 100\,GbE\newline $1\times$ FDR IB &
$2\times$ NVIDIA RTX PRO 6000 Blackwell \newline $1\times$ Xilinx Alveo U55C (in one node) \\
\hline
\addlinespace[4pt]
\multicolumn{7}{l}{\textbf{ARGOS (interferometric pathfinder array)}} \\
\addlinespace[4pt]
Processing nodes &
3 &
$2\times$ AMD EPYC 9354 &
768\,GB &
$8\times$ 16\,TB HDD &
$3\times$ dual 100\,GbE &
$2\times$ NVIDIA L40s \\
\hline
\end{tabularx}
\end{table*}

\subsection{SKAMPI: A Robotic Telescope Built Around the EDD}
The SKAMPI radio telescope is situated at the site of the future Square
Kilometre Array (SKA) in South Africa and has been built as a
prototype and technology demonstrator according to SKA-mid specifications.
Primarily intended for technology commissioning, SKAMPI also supports
independent science programmes. It is equipped with two cryogenically cooled
receivers, one operating in the S-band from \SIrange{1.75}{3.75}{\giga\hertz}
and one with \SI{2}{\giga\hertz} instantaneous bandwidth within
\SIrange{11}{18}{\giga\hertz} in the Ku-band.

Data from both receivers is fed into an EDD backend instance currently running
spectropolarimetry~\citep{Juenemann2026}, pulsar timing, and
VLBI~\citep{Jompoj2026} observations. The SKAMPI EDD backend has been
operational since November 2020 and has received continuous updates including a
full replacement of the computing hardware. The current cluster hardware is
summarised in Table~\ref{tab:hardware}. Unlike the disaggregated architecture
at Effelsberg, SKAMPI uses a converged design in which processing and storage
services share a small number of servers with dedicated fabrics for data
acquisition and storage. For SKAMPI we developed a robotic telescope control
system (TCS) that tightly integrates into the EDD and follows similar design
principles. The TCS runs alongside the backend on the same hardware, reusing
the EDD databases, monitoring infrastructure, and container orchestration. It
is designed for autonomous operation with minimal human supervision, handling
observation scheduling, initial post processing, data quality control, and
transfer of data products to the long-term archive at MPIfR in Bonn as well as
project specific destination. The SKAMPI deployment demonstrates that the EDD
can serve not only as a backend added to an existing telescope, but as the
foundation around which a complete observation system -- including telescope
control -- can be constructed.

\subsection{ARGOS: An Interferometric Prototype Array} ARGOS is a European
Commission Horizon 2020 project undertaking a conceptual design study for a
low-cost, large collecting area radio array targeting more than 1000 antennas
operating in the \SIrange{1}{3}{\giga\hertz} band. The design
incorporates novel elements, including 3D-printed metallised feed horns. A
prototype array is currently being commissioned on the FORTH campus in
Heraklion, Crete, consisting of five 6-m antennas equipped with
dual-polarisation feeds.

The ARGOS prototype backend differs from the other EDD deployments in several
respects. At the frontend, ARGOS employs an RFSoC-based digitiser and
F-engine~\citep{men2025argos}, rather than the EDD frontend system used at
Effelsberg, SKAMPI, and the TNRT. The RFSoC performs digitisation,
channelisation, and delay tracking for the interferometer, delivering
channelised voltage data directly to the EDD backend over 100-GbE. The EDD
backend operates on these data streams without requiring modifications to its
core architecture.

The backend is deployed on a three-node hyper-converged cluster in which
compute, storage, and networking share a single 100-GbE fabric, in contrast to
the disaggregated architecture at Effelsberg and the converged design at
SKAMPI. As at Effelsberg, the system is virtualised using a KVM-based
hypervisor, allowing the backend hardware to be shared with the observatory's
science processing system, which performs interferometric imaging and
calibration. The node specifications are summarised in
Table~\ref{tab:hardware}. A CephFS distributed filesystem provides shared
storage across the cluster.

The central processing component at ARGOS is a GPU-accelerated Tensor Core
correlator-beamformer implemented as part of a general correlation and
beamforming
plugin\footnote{\url{https://gitlab.mpcdf.mpg.de/mpifr-bdg/edd/cbf}} for the
EDD. This plugin and the ARGOS backend will be described in a dedicated future
publication. The ARGOS telescope control
system interfaces directly with the EDD via KATCP, without requiring an
intermediate translation layer, as the control system was designed around this
protocol.

The ARGOS deployment extends the application of the EDD backend from
single-dish systems to interferometric arrays. Current commissioning activities
include real-time correlation and beamforming, commensal transient searching,
pulsar timing, and VLBI recording. These tests are intended to inform the
design of backend systems for larger-scale deployments within the ARGOS
framework.

\vspace{0.5cm} Across these deployments, the EDD backend operates over a wide
range of system scales, from small two-node installations to large multi-tier
clusters, and supports both single-dish and interferometric observing modes.
The underlying hardware architectures differ substantially, including
disaggregated, converged, and hyper-converged configurations, as well as
variations in frontend digitisation and data transport. Despite these
differences, all systems are realised using a common software framework, run the same binary code, with
site-specific adaptations limited to configuration and plugin composition. This
reflects the separation between core architecture and extensible components
described in Section~\ref{sec:system_architecture}, and illustrates the
applicability of the EDD backend across heterogeneous observatory environments.

\section{Conclusions}

The EDD backend addresses an ongoing challenge in modern radio astronomy: wide-band, directly digitised receivers increasingly require backend systems that can be reconfigured, maintained, and extended without rebuilding the instrument software for each observing mode or telescope. We have described a backend framework in which common services for control, deployment, monitoring, and data transport are separated from both site-specific configuration and observing-mode-specific processing pipelines.

The deployments presented here show that this separation is effective in practice. The same framework supports operations on both single dishes and interferometers, while accommodating substantially different hardware, network, and control-system environments. Using Effelsberg pulsar timing data, we have shown that EDD provides performance comparable to prior instrumentation when both systems process the same analogue signal path. When combined with in-receiver digitisation, however, EDD enables significant improvements in sensitivity and timing precision by avoiding degradation in the analogue transmission path.

EDD is released as open-source software to make these backend components, deployment tools, and observing-mode abstractions available beyond the installations described in this paper. In this form, it provides a community resource for developing and operating maintainable real-time backends for broadband radio astronomy instruments.
\begin{acknowledgements}

Partly based on observations with the 100-m telescope of the MPIfR (Max-Planck-Institut f\"{u}r Radioastronomie) at Effelsberg. The 40-m Thai National Radio Telescope is based on the national flagship project for human capacity building and technology development through radio astronomy and geodesy in Thailand, which is supported by Ministry of Higher Education, Science, Research and Innovation (MHESI) for the telescope construction and the development, upgrade, and installation of receivers. SKAMPI, the SKA-MPG prototype telescope, is a facility of the Max-Planck Society (MPG) and was established with the assistance of the South African Radio Observatory (SARAO). It is jointly operated and maintained by the Max Planck Institute for Radio Astronomy (MPIfR) and SARAO. This research was made possible with the support of the MPIfR and SARAO. This project has received funding from the MPG and the European Union's Horizon Europe research and innovation programme under grant agreements No 101093934 (RADIOBLOCKS) and 101094354 (ARGOS-CDS).

\end{acknowledgements}

\bibliographystyle{aa}
\bibliography{biblio}

\end{document}